\documentclass[reprint,amsmath]{revtex4-1}

\usepackage{graphicx}

\begin{document}


\title{Stability of transverse dunes against perturbations; a theoretical study using dune skeleton model}


\author{Hirofumi Niiya}
\author{Akinori Awazu}
\author{Hiraku Nishimori}
\affiliation{Department of Mathematical and Life Sciences, Hiroshima University, Higashihiroshima, Hiroshima 739-8526, Japan}




\date{\today}

\begin{abstract}
The {\it dune skeleton model} is a reduced model to describe the formation process and dynamics of characteristic types of dunes emerging under unidirectional steady wind.
Using this model, we study the dependency of the morphodynamics of transverse dunes on the initial random perturbations and the lateral field size.
It was found that
i)
an increase of the lateral field size destabilizes the transverse dune to cause deformation of a barchan,
ii)
the initial random perturbations decay with time by the power function until a certain time; thereafter, the dune shapes change into three phases according to the amount of sand and sand diffusion coefficient,
iii)
the duration time, until the transverse dune is broken, increases exponentially with increasing the amount of sand and sand diffusion coefficient.
Moreover,
under the condition without the sand supply from windward ground,
the destabilization of transverse dune in this model qualitatively corresponds to the subaqueous dunes in water tank experiments.
\end{abstract}


\maketitle

\section{Introduction}
Erosion due to wind sculpts deserts on Earth and surfaces on Mars, Titan into sand dunes such as barchan, transverse, longitudinal, star-shaped, and dome-shaped
\citep{mckee1979,cooke1993,bourke2009,rubin2009,bridges2012}.
As the dominant factors dictating several dune shapes,
the steadiness of wind direction and the amount of available sand in each dune field are known
\citep{livingstone1996}.
For example,
a unidirectional wind generates barchans of a crescent-shaped or transverse dunes extending perpendicular to the wind direction depending on the amount of available sand,
whereas a bidirectional wind generates longitudinal dunes extending parallel to the sum of two wind directional vectors.
Additionally, a multidirectional wind generates star dunes consisting of multiple crest lines extending from the top to several directions.
As one of remarkable subject in dune studies,
the stability of transverse dunes has remained to be an open issue for both geomorphology and geophysics.
In rescaled water tank experiments, an isolated transverse dune was shown unstable and deformed into a set of barchans as the end-stage
\citep{hersen2002,endo2005,groh2008,reffet2010}.
Also, computer models have reproduced the qualitative and quantitative morphodynamics similar to subaqueous dunes
\citep{nishimori1993,nishimori1998,werner1995,kroy2002,duran2005,zhang2010,katsuki2011}.
Recently, theoretical approaches for the stability of transverse dunes have been conducted by the methodology considered the mass conservation of sand and the sand flows on dunes
\citep{niiya2010,niiya2012,parteli2011,melo2012}.
Especially, Niiya et al. consider the dynamics of a dune as a combination of two-dimensional cross sections (hereafter, 2D-CSs) parallel to the wind direction and proposed the {\it dune skeleton model} (hereafter, {\it DS model}) consisting of coupled ordinary differential equations.
This model is based on several assumptions used in the previous analytical model named ``aeolian/aqueous barchan collision dynamical equations (ABCDE)'' by Katsuki and Nishimori.
ABCDE is able to describe the collision dynamics of two 3D barchans focusing on the central 2D-CS of barchans
\citep{katsuki2005,nishimori2009}.
So far, the {\it DS model} has successfully reproduced three typical shapes of dunes, straight transverse dune, wavy transverse dune, and barchan, depending on the amount of available sand and wind strength
\citep{niiya2010}.
Moreover, the {\it reduced DS model}, which is a further simplified {\it DS model} with a two-variable ordinary differential equation, enabled the elucidation of the mechanism of transition between different dune shapes using a bifurcation analysis
\citep{niiya2012}.
However, in these previous studies, the initial condition was given as a sinusoidal curve with small amplitude and single-wavelength.
Thus, the stability of dunes against random perturbations is yet to be identified.
In this study, we investigate the effect of initial perturbations on the stability of transverse dunes, and the field size effect on the stability is also studied.

\section{Dune skeleton model}
\begin{figure}[t]
\begin{center}
\includegraphics[width = 3.4 in]{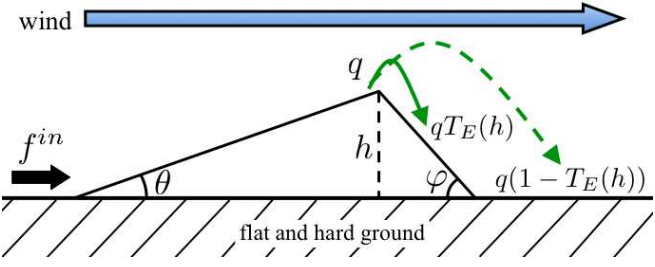}
\caption{
Intra-2D-CS sand flow.
The over-crest sand flux $q$ and {\it sand trapping efficiency} $T_E$ govern the intra-2D-CS flow.
The green solid and dashed lines indicate sand deposition on the downwind slope and escaping sand to the leeward ground, respectively.
}
\label{fig:intra-2D-CS}
\end{center}
\end{figure}

The {\it DS model} covers the formation processes of barchans and transverse dunes, both of which are generated under a unidirectional steady wind.
This model is roughly based on three considerations.
First, the dunes consist of 2D-CSs as mentioned in the previous section.
Second, a lateral distance between neighboring 2D-CSs is set constant.
Third, a combination of two forms of sand movement, intra-sand movement within each 2D-CS and inter-sand movement between neighboring 2D-CSs, is considered to govern the macroscopic morphodynamics of dunes.
3D barchans are isolated on hard ground in both wind and lateral directions,
whereas 3D transverse dunes are not isolated because they extend in the lateral direction.
However, in the wind direction, the laterally extending 3D transverse dunes are assumed to be separated by inter-dune hard ground.
In addition, considering the observation that the 2D-CSs of barchans and transverse dunes very roughly show a scale-invariant triangular shape, we assume that the angles of their upwind and downwind slopes ($\theta$ and $\varphi$, respectively as shown in Fig. \ref{fig:intra-2D-CS}) are maintained constant through their migration, irrespective of their size.
From these assumptions, the horizontal (i.e., wind directional) position and the height of each 2D-CS are uniquely determined if only the coordinate ($x,h$) of its crest is given.
Moreover, the empirical geometrical constants A, B, and C are introduced as
\begin{eqnarray*}
{\rm A}=\frac{\tan{\theta} \tan{\varphi}}{\tan{\theta} + \tan{\varphi}},
{\rm B}=\frac{\tan{\varphi}}{\tan{\theta} + \tan{\varphi}},
{\rm C}=\frac{\tan{\theta}}{\tan{\theta} + \tan{\varphi}},
\end{eqnarray*}
where A, B, and C are set to 1/10, 4/5, and 1/5, respectively, reflecting the typical 2D-CS profiles of real barchans and transverse dunes.
As mentioned above, sand flow is classified into two forms:
(a) the intra-2D-CS flow and (b) the inter-2D-CS flow.
The intra-2D-CS flow along the upwind slope is uniquely determined if the over-crest sand flux $q$ and the incoming sand flux from the windward ground $f^{in}$ are given
(Fig. \ref{fig:intra-2D-CS}).
The over-crest sand flux $q$ determines the erosion rate of the 2D-CS's upwind slope and the over-crest blown sand deposits along the downwind slope or directly escapes to the leeward inter-dune ground.
The deposition ratio $T_E$ of the over-crest sand along the downwind slope is assumed as an increasing function of height of the specific form
\begin{eqnarray}
\label{eq:sand-trapping-efficiency}
T_E(h)=\frac{h}{1.0 + h}.
\end{eqnarray}
Note that $T_E(h)$ is termed as the {\it sand trapping efficiency} \citep{momiji2000} and Eq. (\ref{eq:sand-trapping-efficiency}) roughly represents the case of typical shear velocity $u_*=0.4$ m\hspace{0.5 ex}s$^{-1}$ at the dune crest, where $h$ is the height in a meter unit.
The quantity of $q$ reflects the over-dune wind strength; here, we assume $q$, where $0.1 \le q \le 1.0$, to be constant in each desert field, independent of the 2D-CS's height.
In addition, all of the incoming sand flux from the windward ground, $f^{in}$, deposits on the upwind slope of dunes.

\begin{figure}[t]
\begin{center}
\includegraphics[width = 3.4 in]{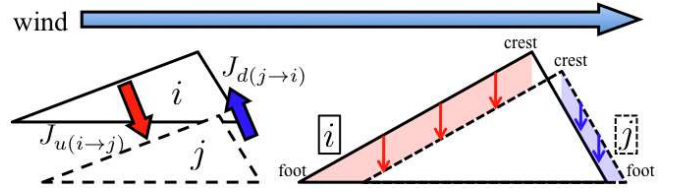}
\caption{
Inter-2D-CS sand flow $J_u$ (upwind) and $J_d$ (downwind).
Both the red and blue arrows indicate local sand transport depending on the height difference.
The red and blue areas correspond to $J_u$ and $J_d$, respectively, by summing the local sand transport from the slope's foot to the 2D-CS's crest.
}
\label{fig:inter-2D-CS}
\end{center}
\end{figure}

The inter-2D-CS flow $J_{u(i\to j)}/J_{d(j\to i)}$ occurs only between the upwind/downwind slopes of the neighboring 2D-CSs, $i$ and $j$
(Fig. \ref{fig:inter-2D-CS}).
Locally, most of the lateral sand transport is determined by the height difference between neighboring 2D-CS's slopes.
Therefore, we assume that the total flux is the sum of local sand transport from the slope's foot to the 2D-CS's crest.
Namely, the flux is roughly considered as the lateral diffusion depending on the height difference, though the consideration of the overlap length of slopes causes a nonlinearity in the present form of the inter-2D-CS flux.
The specific forms of $J_{u(i\to j)}$ and $J_{d(j\to i)}$ are
{\small
\begin{subequations}
\begin{eqnarray}
\nonumber
J_{u(i\to j)}=
\left\{
\begin{array}{lr}
 \frac{D_u{\rm B}}{{\rm 2A}\Delta w^2}\left\{h_i^2-\left[h_j-\frac{\rm A}{\rm B}(x_j-x_i)\right]^2\right\} & x_j-x_i > 0\\
 \frac{D_u{\rm B}}{{\rm 2A}\Delta w^2}\left\{\left[h_i+\frac{\rm A}{\rm B}(x_j-x_i)\right]^2-h_j^2\right\} & x_j-x_i \le 0
\end{array}
\right.\\
\label{eq:inter-up}
\\
\nonumber
J_{d(j\to i)}=
\left\{
\begin{array}{lr}
 \frac{D_d{\rm C}}{{\rm 2A}\Delta w^2}\left\{h_j^2-\left[h_i-\frac{\rm A}{\rm C}(x_j-x_i)\right]^2\right\} & x_j-x_i > 0\\
 \frac{D_d{\rm C}}{{\rm 2A}\Delta w^2}\left\{\left[h_j+\frac{\rm A}{\rm C}(x_j-x_i)\right]^2-h_i^2\right\} & x_j-x_i \le 0,
\end{array}
\right.\\
\label{eq:inter-down}
\end{eqnarray}
\end{subequations}
}
where $\Delta w$, set as $\Delta w=1$ hereafter, is the lateral interval between neighboring 2D-CSs.
These quantities correspond to the colored areas in Fig. \ref{fig:inter-2D-CS} multiplied by diffusion coefficients.
Here, the upwind and downwind diffusion coefficients ($D_u$ and $D_d$, respectively) control the amount of inter-2D-CS sand flow on the respective sides of the slopes and reflect the over-dune wind strength.
With consideration of the above intra- and inter-sand flows, the dynamics of the coordinates $(x_i,h_i)\hspace{0.5 ex}(i=1,\cdots,N)$ of the 2D-CS's crest are given as a system of coupled ordinary differential equations
\citep{niiya2010,niiya2012}:
{\small
\begin{subequations}
\begin{align}
\label{eq:dsma}
\frac{dx_i}{dt}=&\frac{1}{h_i}\left[q({\rm B}T_E(h_i)+{\rm C})+\sum_{j=i\pm1}({\rm B}J_{d(j\to i)}+{\rm C}J_{u(i\to j)})-{\rm C}f_i^{in}\right],\\
\label{eq:dsmb}
\frac{dh_i}{dt}=&\frac{\rm A}{h_i}\left[q(T_E(h_i)-1)+\sum_{j=i\pm 1}(J_{d(j\to i)}-J_{u(i\to j)})+f_i^{in}\right].
\end{align}
\end{subequations}
}
\begin{figure}[t]
\begin{center}
\includegraphics[width = 3.4 in]{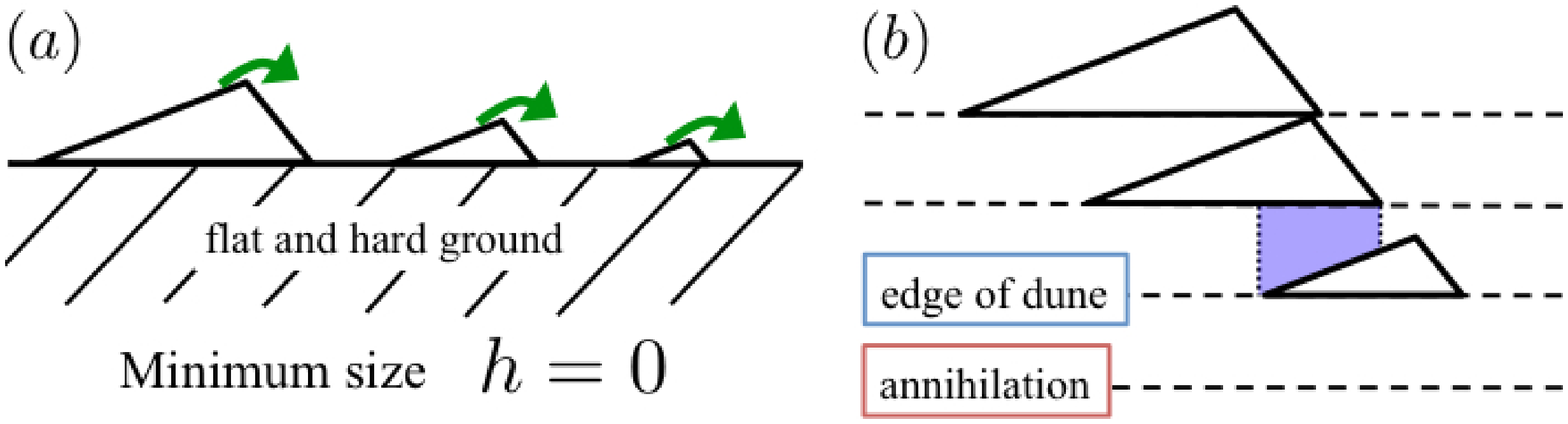}
\caption{
Schematic of annihilation rule.
(a) Annihilation by escaping of sand from 2D-CS.
$h=0$ mean the exposed state of ground.
(b) Annihilation by leaving 2D-CS from dune.
The blue area is overlap between the 2D-CS at the edge and its neighbor.
}
\label{fig:annihilation}
\end{center}
\end{figure}
We also introduce the annihilation rule of the 2D-CSs at the lateral edges of a dune;
this rule is required to simulate the shrinking process of dunes.
This rule is applied in the cases
where $h_i$ is lower than $0$ by the escaping of sand from $i$-th 2D-CS (Fig. \ref{fig:annihilation}(a))
or where the overlap between the 2D-CS at the edge and its neighbor vanishes (Fig. \ref{fig:annihilation}(b)).
If the annihilation rule is applied to $i$-th 2D-CS, we consider that the state of $i$-th lane gets corresponding to the exposed state of ground, that is,
$i$-th 2D-CS is taken out of the calculation though the virtual crest of height zero is fixed at the foot of downwind slope as
\begin{eqnarray*}
h_i=0,\hspace{2 ex}x_i=\max_{j=i\pm 1}\left[x_j+\frac{\rm C}{\rm A}h_j\right].
\end{eqnarray*}
In order to compare the results obtained from {\it DS model} with the real desert dunes,
we estimate the value of $q$, $D_u$, and $D_d$ in physical units on the basis of the observed data in real dune fields.
The migration velocity of single 2D-CS in our model is assumed as
\begin{eqnarray}
\label{eq:single-velocity}
\frac{dx}{dt}=\frac{q}{h}T_E(h)=\frac{q}{1+h}.
\end{eqnarray}
Indeed, in field observations
\citep{finkel1959,hastenrath1967,hastenrath1987},
the migration speed $c$ (m/year) and size $H$ (m) of barchans are reasonably fitted as \citep{andreotti2002}
\begin{eqnarray}
\label{eq:fitted-velocity}
c\simeq \frac{Q}{H_0+H}=\frac{Q/H_0}{1+H/H_0},
\end{eqnarray}
where $Q$ is the volumic sand flux (m$^2$/year), $H_0$ is the cut-off height ranging from $1.8$ m (Finkel 1955-58) to $10.9$ m (Hastenrath 1958-64).
Comparing Eq. (\ref{eq:single-velocity}) with Eq. (\ref{eq:fitted-velocity}), the relationship between our model and real data is given as
\begin{eqnarray*}
h=\frac{H}{H_0}, q=\frac{Q}{H_0}=c\left(1+\frac{H}{H_0}\right).
\end{eqnarray*}
Here, we roughly assume that a dune with $5$ m height migrates $10$ m per year and the cut-off height is $1$ m, namely, $c=10,H=5$, and $H_0=1$ in meter-year unit.
Converting time unit from year to day, we get about $q\approx 0.164$ m$^2$/day.
Also, $q$ used in our model takes the value from $0.1$ to $1.0$.
Therefore, characteristic time unit and space unit accompanied with physical values used in our model are approximately estimated as a day and a meter, respectively.

\section{Results}
Numerical simulation of Eqs. (\ref{eq:dsma}) and (\ref{eq:dsmb}) is carried out using several numbers of 2D-CSs, $N=200$, $600$, $1000$, $1400$, and $1800$.
The lateral boundary condition is set as periodic,
whereas the wind directional boundary condition is set such that sand escaping from the leeward boundary is redistributed uniformly from the windward boundary.
Thus, mass conservation, i.e., the total area of 2D-CSs, is maintained throughout the simulation unless the annihilation of a 2D-CS occurs.
Moreover, according to the rule for the redistribution of escaping sand,
the incoming sand flux to each 2D-CS from the windward ground $f^{in}$ is expressed as
\begin{eqnarray*}
f_{i}^{in}=\frac{1}{N}\sum_{i=1}^{N}q(1-T_E(h_i)),
\end{eqnarray*}
where $T_E$ is the {\it sand trapping efficiency} given by Eq. (\ref{eq:sand-trapping-efficiency}).
In order to elucidate the stability of the transverse dunes against a perturbation,
the initial dune shape is set slightly fluctuated from a straight transverse dune.
More specifically,
while the initial heights of 2D-CS's crests are set uniform, i.e., $h_i(0)=H_0$,
the fluctuation of initial wind directional positions of the crest of each 2D-CS is set random with an amplitude depending on $H_0$; that is, $x_i(0)=\epsilon (H_0)$.
Here, $\epsilon (H_0)$ is a random number uniformly distributed between $0$ and $H_0/10$.
\begin{figure}[t]
\begin{center}
\includegraphics[width= 3.4 in]{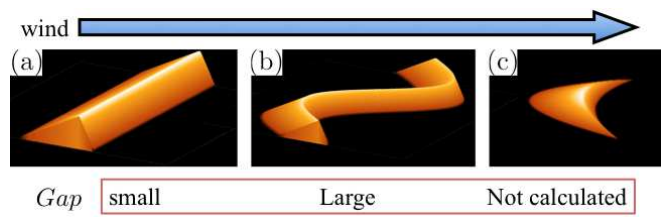}
\caption{
Typical dune shapes depending on $Gap(t)$ by numerical simulation:
(a) straight transverse dune,
(b) wavy transverse dune, and
(c) barchan.
}
\label{fig:dune-shape}
\end{center}
\end{figure}
Like in the previous paper (Niiya et al., 2010),
we classify the obtained dune shapes into three different phases according to the quantity,
\begin{eqnarray}
\label{eq:gap}
Gap(t)=\frac{1}{N}\sum_{i=1}^{N}(x_i(t)-x_{i+1}(t))^2.
\end{eqnarray}
$Gap$ describes the degree of sinuosity for a transverse dune
(Fig. \ref{fig:dune-shape}).
\begin{description}
\item[I)]
If $Gap(10^8)/Gap(0)<10^{-2}$ is satisfied,
we regard a straight transverse dune is formed.
The phase in this state is referred to as the ``ST phase''.
\item[II)]
If $Gap(10^8)/Gap(0)\ge10^{-2}$ is satisfied,
we regard a wavy transverse dune is formed.
The phase in this state is referred to as the ``WT phase''.
\item[III)]
If the annihilation of a 2D-CS occurs before $t=10^8$,
we regard the wavy transverse dune deforms to the barchan.
The phase in this state is referred to as the ``B phase''.
\end{description}
Note that, through an analytical approach of {\it reduced DS model} with only two 2D-CSs,
the transverse dunes are shown to be attracted to one of ST, WT, and B phases
\citep{niiya2012}.
In this simulation of the {\it DS model},
we fix the over-crest sand flux in each 2D-CS, $q=0.5$ (m$^2$/day), which reflects the wind strength in a dune field,
downwind diffusion constant given in Eq. (\ref{eq:inter-down}) is set $D_d=0.1$ (m$^2$/day), and the lateral period between neighboring 2D-CSs is assumed as $1$ m.
Then, three parameters, $N$, $H_0$, and $D_u$ are varied,
where $N$ (m) is the number of 2D-CSs, i.e., the lateral field size in meter,
$H_0$ (m) is the uniform initial height in meter of 2D-CSs, which controls the total amount of sand in the system, and
$D_u$ (m$^2$/day) is the upwind diffusion coefficient introduced in Eq. (\ref{eq:inter-up}).
The numerical calculations are performed until the simulation time reaches $t=10^{8}$ (day) or the annihilation of a 2D-CS occurs.
If annihilation occurs, a dune rapidly shrinks and eventually disappears.
We take five ensembles of calculation with different amplitude of initial perturbations for each set of parameters.
Obtained results are as follows.
\begin{figure}[t]
\begin{center}
\includegraphics[width= 3.4 in]{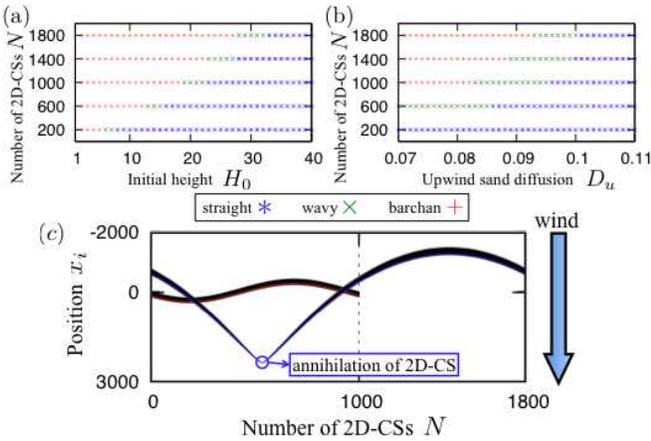}
\caption{
Phase diagram of dune shapes obtained from numerical simulations by varying the two pairs of control parameters:
(a) ($H_0,N$) and (b) ($D_u,N$).
The blue, green, and red marks denote the ST, WT, and B phase independent of the initial random perturbation, respectively.
Other parameters are set as (a) $D_u=0.1$ and (b) $H_0=30$.
(c) Typical shape of transverse dune against field size: (red) $N=1000$ and (blue) $N=1800$.
The dune with red crest line represents the stationary wavy transverse dune in WT phase,
whereas the dune with blue one represents the dune at the moment when the annihilation of a 2D-CS occurs.
The parameters are fixed as ($H_0,D_u$)=($20,0.1$).
}
\label{fig:N-result}
\end{center}
\end{figure}
First, according to the above classification rules {\bf I)}$\sim${\bf III)},
we verify the influence of the lateral field size on the final dune phase.
In Figs. \ref{fig:N-result}(a) and \ref{fig:N-result}(b), the final dune phases obtained in two sets of parameter spaces $(H_0,N),(D_u,N)$ are classified.
Figure \ref{fig:N-result}(c) shows the lateral system size dependency of the stability of transverse dunes,
where initially same height ($H_0=20$) of transverse dunes are set in different lateral size ($N=1000,1800$) of fields.
The transverse dune in a longer lateral size of field will break,
whereas the transverse dune in a shorter lateral size of field is kept steadily in the WT phase.
It means that the critical wave length $\lambda_c$ for the occurrence of the instability of transverse dune exists between $1000$ m and $1800$ m in this case.
It should be noted that $\lambda_c$ depends also on $D_u$ and $D_d$,
so it is required to estimate the specific values of them in respective dune fields,
which is the next issue.
Another relevant factor for the stability of a transverse dune in the present model is the effect of homogeneous redistribution of sand from the upper boundary,
that seems to stabilize a transverse dune through supplying a same amount of sand at the upwind slopes of both the thin-and-low part and the thick-and-high part of a transverse dune,
the slowing down effect for the former part is more obvious than that for the latter part.
Therefore, the destabilization by the fasten migration of the lower part is suppressed.
\begin{figure}[t]
\begin{center}
\includegraphics[width= 3.4 in]{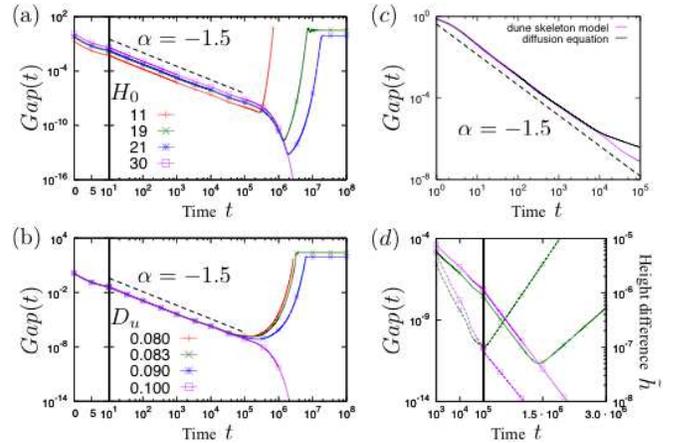}
\caption{
Typical time evolutions of dune sinuosity using $Gap$: (a) $H_0$ and (b) $D_u$.
The solid lines are selected from the simulations with $N=1000$ in Fig. \ref{fig:N-result}(a).
$\alpha$ is the power index.
(c) Comparison {\it DS model} and one-dimensional diffusion equation until $t=10^5$.
(d) Comparison $Gap$ (solid) and height difference $\tilde{h}$ (dashed) in the system.
The lines are picked from (a): (green) $H_0=19$ and (purple) $H_0=30$.
}
\label{fig:gap-result}
\end{center}
\end{figure}
Second, in order to see the evolution process of transverse dunes with a fixed lateral size ($N=1000$), we focus on the time evolution of $Gap$.
Specifically,
we pick up four points from Fig. \ref{fig:N-result}(a):
$(H_0,N)=(11,1000)$ in B phase (named point-a),
$(H_0,N)=(19,1000)$ in WT phase near the boundary to B phase (named point-b),
$(H_0,N)=(21,1000)$ in WT phase near the boundary to ST phase (named point-c),
and $(H_0,N)=(30,1000)$  in ST phase (named point-d).
Figures \ref{fig:gap-result}(a) and \ref{fig:gap-result}(b) show the time evolution of $Gap$, and it is characteristic that
the initial fluctuation of all above defined points decays with time in the power of $\alpha \simeq -1.5$ until $t=10^5$; that is, the initial transverse dunes temporarily approach the straight shape of transverse dune.
Thereafter, $Gap$ at point-a turns into a drastic increase, where the annihilation of a 2D-CS takes place, namely, the breaking of a transverse dune occurs.
In contrast, $Gap$ at the remaining three points, b, c and d, once increases the decreasing rate, among which $Gap$ at point-b and point-c turn into increasing after certain durations of time, and finally converge to the respective final states.
Especially, at point-b, $Gap$ oscillates with a small amplitude in the final state,
whereas $Gap$ settles to a constant value at point-c.
To understand these characteristic evolution of $Gap$, we divide the simulation period into two parts; i) power-law decaying time region until $t=10^5$ and ii) drastic growing or decaying time region after $t=10^5$.
For i), remember that the perturbation $\delta x_i$ added to the initial straight transverse dune with $x_i=x_0, h_i=H_0$ was randomly selected value between $0$ and $H_0/10$.
Here, we assume that
the power decay of $Gap$ corresponds to the smoothing process of the initial perturbation caused by the lateral sand transport.
To check this, the one-dimensional diffusion equation
\begin{eqnarray}
\label{eq:diffusion}
\frac{\partial f_i}{\partial t}=D(f_{i+1}+f_{i-1}-2f_i)\hspace{2 ex} (i=1,2,\cdots,L)
\end{eqnarray}
under the same periodic condition as the {\it DS model} is numerically calculated,
where the variables in Eq. (\ref{eq:diffusion}) correspond as $(f_i,L,D) \approx (x_i,N,D_u)$ with $f_i(0)=f(0)+\delta f_i\hspace{1 ex} (\delta f_i=\delta x_i)$,
and the time evolution of $Gap(t)=\sum_{i=1}^{N}(f_i(t)-f_{i+1}(t))^2$ is measured.
Then, the decay rate of $Gap$ in the discrete diffusion equation (Eq. (\ref{eq:diffusion})) is confirmed to obey the power-law with exponent $\alpha \simeq-1.5$.
This exponent is analytically derived from a simple dimension analysis of the diffusion equation (Fig. \ref{fig:gap-result}(c)).
In this way, the power-law decay of $Gap$ in the {\it DS model} in the time region i) is strongly guessed to originate from the lateral diffusion of sand.
For ii),
the time evolution of $Gap$ after $t=10^5$ shows the exponential growth or decay (Fig. \ref{fig:gap-result}(d)),
where the amplitude of single wave shown in Fig. \ref{fig:N-result}(c) increases or decreases.
In order to see this drastic change of the time evolution of $Gap$ from another aspect,
we measure the difference between the maximum and minimum heights of 2D-CSs, $\tilde{h}=h_{max}-h_{min}$.
Here, $Gap$ increases following the increase of $\tilde{h}$ because the velocity of 2D-CSs is inversely proportional to their height from Eq. (\ref{eq:dsma}).
Actually, in Fig. \ref{fig:gap-result}(d), $\tilde{h}$ turns into increasing before the increase in $Gap$.
Thus, the transition of $Gap$ from decay to growth is induced by the increase in height difference of 2D-CSs.
In the next, we consider the mechanism of the exponential increase or decrease in $Gap$.
In the diffusion equation (Eq. (\ref{eq:diffusion})),
if the wave number $k$ of transverse dune is determined,
$f_i(t)$ decay exponentially with time according to the decay rate $-Dk^2$.
In contrast, the {\it DS model} consists of the nonlinear coupling of position and height of 2D-CSs as seen in Eqs. (\ref{eq:dsma}) and (\ref{eq:dsmb}),
which may act as the factor to remain the wavy transverse dune stable.
Namely,
the exponential decay of $Gap$ is generated by the diffusion,
whereas the exponential growth comes from the coupling between position and height.
This problem is left as the next issue.
\begin{figure}[t]
\begin{center}
\includegraphics[width= 3.4 in]{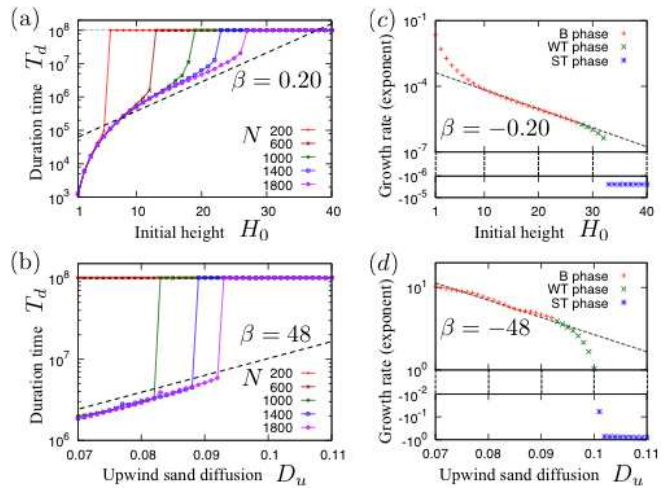}
\caption{
Duration times of transverse dunes and growth rate of $Gap$ after diffusion process: (a, c) $H_0$ and (b, d) $D_u$.
$\beta$ is the exponent.
(a, b) The colors of the points indicate the different $N$ used for the simulations.
Each point is the average duration time per simulation.
(c, d) The fitting values by exponential function.
$N$ is fixed as $N=1800$.
}
\label{fig:duration-result}
\end{center}
\end{figure}
Finally, we study the duration time $T_d$, which represents the time interval of transverse dunes before the break of the crestline.
Note that the maximum value of $T_d$ equals the simulation time $t=10^8$.
Figures \ref{fig:duration-result}(a) and \ref{fig:duration-result}(b) show that the values $T_d$ lower than $10^8$ obtained in the B phase is nicely fitted by exponential function of two control parameters as $T_d\propto \exp(0.2H_0+48D_u)$,
though an increase in $N$ extends the range of the B phase.
To elucidate the cause of these exponential increases of $T_d$ as functions of $H_0$ and $D_u$,
we focus on the exponential growth region of $Gap$ after diffusion process in Fig. \ref{fig:gap-result},
where $Gap(t)\propto \exp(\sigma t)$ holds with growth rate $\sigma$.
If $Gap$ exceeds a certain value and the transverse dune breaks,
therefore, the duration time is roughly estimated as $T_d\propto 1/\sigma$.
We measure $\sigma$ from simulation data as shown in Figs. \ref{fig:duration-result}(c) and \ref{fig:duration-result}(d) and $\sigma$ in the B phase is nicely fitted as $\sigma \propto \exp(-0.2H_0-48D_u)$, which relations correspond to the dependency of $T_d$ on $H_0$ and $D_u$ shown in Figs. \ref{fig:duration-result}(a) and \ref{fig:duration-result}(b).

\section{Conclusion}
\begin{figure}[t]
\begin{center}
\includegraphics[width = 3.4 in]{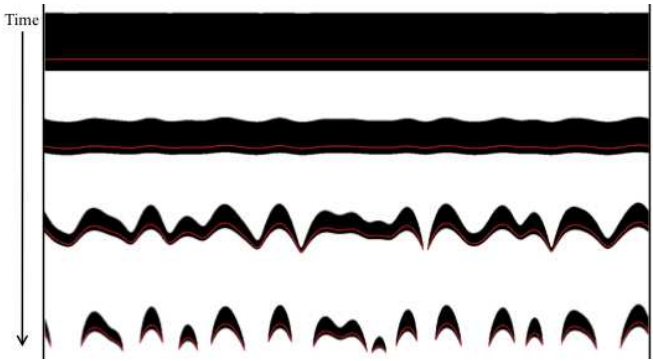}
\caption{
Deformation of transverse dune without redistribution rule of sand.
}
\label{fig:barchanoid}
\end{center}
\end{figure}

In this study, we discussed on the stability of transverse dunes against random perturbations and its dependency on lateral size using the {\it DS model}.
Numerical simulations show that
i) an increase in the number of 2D-CSs, namely, the lateral size increase, destabilizes the transverse dunes,
ii) in the initial stage of simulations, the small scale of fluctuation of a transverse dune is smoothed by the lateral sand diffusion, and
iii) after the initial diffusion regime of time, transverse dunes evolve according to the specific growth rate depending on two control parameters, $H_0$ and $D_u$.
The uniform redistribution rule of sand in the present model affects on the stability of transverse dunes.
Specifically, the escaping sand from a transverse dunes is uniformly supplied from the windward boundary in the present model.
Note that most of previous water tank experiments and mathematical models have indicated that the inevitable instability of transverse dunes causes the deformation of them into barchans.
In such cases, the boundary condition in wind direction was set open, namely, no-sand was supplied from the windward ground.
This boundary condition is definitely different from that in the present {\it DS model}.
If the open boundary is used in the present {\it DS model},
the total amount of sand consistently decreases and the transverse dune can not keep the stable states.
To demonstrate this,
we conduct the numerical simulation without the redistribution rule of sand under the same initial condition as the above mentioned case.
The initial condition is randomly given as with this paper.
Then, the initially straight transverse dune deforms into the wavy shaped dune such like barchanoid,
and further simulation causes the emergence of many barchans,
even though the barchans eventually disappear (Fig. \ref{fig:barchanoid}).
In this way, under the no-sand supply condition, our model shows the instability of transverse dunes similar to previous experiments and models.
Based on the this result,
it is reasonably guessed that the morphodynamics of dunes vary largely depending on the supplying condition of sand.

\begin{acknowledgments}
This work was partially supported by the 21st Century COE Program, ``Toward a new basic science depth and synthesis.''
\end{acknowledgments}


\begin{thebibliography}{99}
\bibitem[McKee, 1979]{mckee1979} McKee, E.D., 1979. Introduction to a study of global sand seas. U.S. GPO, Washington, D.C.
\bibitem[Cooke et al., 1993]{cooke1993} Cooke, R.U., Warren, A., Goudie, A., 1993. Desert geomorphology. CRC.
\bibitem[Bourke et al., 2009]{bourke2009} Bourke, M.C., Goudie, A.S., 2009. Varieties of barchan form in the Namib Desert and on Mars. Aeolian Research 1, 45.
\bibitem[Rubin et al., 2009]{rubin2009} Rubin, D.M., Hesp, P.A., 2009. Multiple origins of linear dunes on Earth and Titan. Nature Geoscience 2, 653.
\bibitem[Bridges et al., 2012]{bridges2012} Bridges, N.T., Ayoub, F., Avouac, J-P., Leprince, S., Lucas, A., Mattson, S., 2012. Earth-like sand fluxes on Mars. Nature 485, 339.

\bibitem[Livingstone et al., 1996]{livingstone1996} Livingstone, I., Warren, A., 1996. Aeolian Geomorphology. Addison Wesley Longman, Harlow, United Kingdom.

\bibitem[Hersen et al., 2002]{hersen2002} Hersen, P., Douady, S., Andreotti, B., 2002. Relevant length scale of barchan dunes. Phys. Rev. Lett. 89, 264301.
\bibitem[Endo et al., 2005]{endo2005} Endo, N., Sunamura, T., Takimoto, H., 2005. Barchan ripples under unidirectional water flows in the laboratory: formation and planar morphology. Earth Surf. Proc. Land. 30, 1675--1682.
\bibitem[Groh et al., 2008]{groh2008} Groh, C., Wierschem, A., Aksel, N., Rehberg, I., Kruelle, C.A., 2008. Barchan dunes in two dimensions: Experimental tests for minimal models. Phys. Rev. E 78, 21304.
\bibitem[Reffet et al., 2010]{reffet2010} Reffet, E., Courrech du Pont, S., Hersen, P., Douady, S., 2010. Formation and stability of transverse and longitudinal sand dunes. Geology 38, 491.

\bibitem[Nishimori et al., 1993]{nishimori1993} Nishimori, H., Ouchi, N., 1993. Formation of ripple patterns and dunes by wind-blown sand. 71, 197--200.
\bibitem[Werner, 1995]{werner1995} Werner, B.T., 1995. Eolian dunes: computer simulations and attractor interpretation. Geology 23, 1107.
\bibitem[Nishimori et al., 1998]{nishimori1998} Nishimori, H., Yamasaki, M., Andersen, K.H., 1998. A simple model for the various pattern dynamics of dunes. Int. J. Mod. Phys. B 12, 257--272.
\bibitem[Kroy et al., 2002]{kroy2002} Kroy, K., Sauermann, G., Herrmann, H.J., 2002. Minimal model for sand dunes. Phys. Rev. Lett. 88, 54301.
\bibitem[Dur\'an et al., 2005]{duran2005} Dur\'an, O., Schw\"ammle, V., Herrmann, H., 2005. Breeding and solitary wave behavior of dunes. Phys. Rev. E 72, 21308.
\bibitem[Zhang et al., 2010]{zhang2010} Zhang, D., Narteau, C., Rozier, O., 2010. Morphodynamics of barchan and transverse dunes using a cellular automaton model. J. Geophys. Res. 115, F03041.
\bibitem[Katsuki et al., 2011]{katsuki2011} Katsuki, A., Kikuchi, M., Nishimori, H., Endo, N., Taniguchi, K., 2011. Cellular model for sand dunes with saltation, avalanche and strong erosion: collisional simulation of barchans. Earth Surf. Proc. Land. 36, 372.

\bibitem[Niiya et al., 2010]{niiya2010} Niiya, H., Awazu, A., Nishimori, H., 2010. Three-Dimensional Dune Skeleton Model as a Coupled Dynamical System of Two-Dimensional Cross Sections. J. Phys. Soc. Jpn. 79, 063002.
\bibitem[Parteli et al., 2011]{parteli2011} Parteli, E.J.R., Andrade Jr, J.S., Herrmann, H.J., 2011. Transverse instability of dunes. Phys. Rev. Lett. 107, 188001.
\bibitem[Melo et al., 2012]{melo2012} Melo, H.P.M., Parteli, E.J.R., Andrade Jr, J.S., Herrmann, H.J., 2012. Linear stability analysis of transverse dunes. Arxiv preprint arXiv: 1202.2842.
\bibitem[Niiya et al., 2012]{niiya2012} Niiya, H., Awazu, A., Nishimori, H., 2012. Bifurcation Analysis of the Transition of Dune Shapes under a Unidirectional Wind. Phys. Rev. Lett. 108, 158001.

\bibitem[Katsuki et al., 2005]{katsuki2005} Katsuki, A., Nishimori, H., Endo, N., Taniguchi, K., 2005. Collision Dynamics of Two Barchan Dunes Simulated Using a Simple Model. J. Phys. Soc. Jpn. 74, 538--541.
\bibitem[Nishimori et al., 2009]{nishimori2009} Nishimori, H., Katsuki, A., Sakamoto, H., 2009. Coupled ODEs Model for the Collision Process of Barchan Dunes. Theor. Appl. Mech. Jpn. 57, 179--184.

\bibitem[Momiji et al., 2000]{momiji2000} Momiji, H., Warren, A., 2000. Relations of sand trapping efficiency and migration speed of transverse dunes to wind velocity. Earth Surf. Proc. Land. 25, 1069--1084.

\bibitem[Finkel, 1959]{finkel1959} Finkel, H.J., 1959. The barchans of southern Peru. J. Geology 67, 614.
\bibitem[Hastenrath, 1967]{hastenrath1967} Hastenrath, S., 1967. The Barchans of the Arequipa Region, Southern Peru. Zeitschrift f\"ur Geomorphologie 11, 300.
\bibitem[Hastenrath, 1987]{hastenrath1987} Hastenrath, S., 1987. The barchan dunes of Southern Peru revisited. Zeitschrift f\"ur Geomorphologie 31-2, 167.

\bibitem[Andreotti et al., 2002]{andreotti2002} Andreotti, B., Claudin, P., Douady, S., 2002. Selection of dune shapes and velocities. Eur. Phys. J. B 28, 321.
\end{thebibliography}
\end{document}